\documentclass[aps,prd,twocolumn,10pt,showpacs,superscriptaddress,nofootinbib,preprintnumbers]{revtex4-2}

\usepackage{amssymb,amsmath,verbatim,mathtools,needspace,enumitem,etoolbox,graphicx,physics,microtype,afterpage,xspace,tabularx,lmodern,multirow}
\usepackage{booktabs}
\usepackage{siunitx}
\usepackage{gensymb}
\usepackage{ulem}  

\usepackage{amsmath}
\usepackage{tikz}
\usepackage{tikz-feynman}
\usepackage{booktabs}
\usepackage{bm}
\usepackage{cancel}
\tikzfeynmanset{
  compat=1.1.0
}
\tikzset{every path/.append style={thick}}
\usepackage{amsfonts}
\usepackage{amssymb}
\usepackage{enumerate}
\usepackage{color}
\usepackage{graphicx}
\usepackage{bm}
\usepackage[colorlinks=true,citecolor=blue,urlcolor=blue]{hyperref}
\hypersetup{colorlinks=true,citecolor=romared,linkcolor=romared,urlcolor=romared}
\definecolor{romared}{RGB}{142,0,28}
\usepackage{afterpage}
\usepackage{multirow}
\usepackage{appendix}
\usepackage[table]{xcolor}
\usepackage{feynmp-auto}
\usepackage{aas_macros}
\usepackage{makecell}
\usepackage{soul}

\usepackage{lipsum}
\usepackage{orcidlink}
\definecolor{tabblue}{RGB}{31, 119, 180}
\definecolor{darkblue}{RGB}{0, 0, 120}
\definecolor{tabred}{RGB}{214, 39, 40}
\definecolor{tabgreen}{RGB}{44, 160, 44}
\definecolor{tabgray}{RGB}{100, 100, 100}
\definecolor{taborange}{RGB}{255, 127, 14}
\definecolor{tabbrown}{RGB}{128, 0, 0}
\definecolor{tabpink}{RGB}{255, 141, 161}
\definecolor{tabpurple}{RGB}{148, 103, 189}
\definecolor{goldenrod}{RGB}{218, 165, 32}
\usepackage{float}
\usepackage{placeins}
\newcommand{\bea}{\begin{eqnarray}}
\newcommand{\eea}{\end{eqnarray}}

\begin{document}
\preprint{UTWI-11-2026}

\title{CMB Limits on the Absorption of Light Vector and Axial-Vector Dark Matter}

\author{Gabriele Montefalcone\,\orcidlink{0000-0002-6794-9064}}
\email{montefalcone@utexas.edu}
\affiliation{Texas Center for Cosmology and Astroparticle Physics, Weinberg Institute for Theoretical Physics, Department of Physics, The University of Texas at Austin, Austin, TX 78712, USA}

\author{Nicola Bellomo\,\orcidlink{0000-0002-4375-705X}}
\affiliation{Texas Center for Cosmology and Astroparticle Physics, Weinberg Institute for Theoretical Physics, Department of Physics, The University of Texas at Austin, Austin, TX 78712, USA}
\affiliation{Dipartimento di Fisica e Astronomia G. Galilei, Universit\`a degli Studi di Padova, Via Marzolo 8,
I-35131 Padova, Italy}
\affiliation{INFN, Sezione di Padova, Via Marzolo 8, I-35131, Padova, Italy}
\affiliation{INAF - Osservatorio Astronomico di Padova, Vicolo dell'Osservatorio 5, I-35122 Padova, Italy}

\author{Kimberly K.~Boddy\,\orcidlink{0000-0003-1928-4667}}
\affiliation{Texas Center for Cosmology and Astroparticle Physics, Weinberg Institute for Theoretical Physics, Department of Physics, The University of Texas at Austin, Austin, TX 78712, USA}

\begin{abstract}
Leptophilic sub-MeV spin-1 dark matter (DM) can be converted into a photon via inelastic scattering with a free electron or absorption by a neutral hydrogen atom in the primordial plasma. We study for the first time the impact of the energy injection resulting from such processes on cosmic microwave background (CMB) anisotropies. We obtain upper limits on the vector and axial-vector DM–electron couplings using {\it Planck} 2018 temperature, polarization, and lensing data for DM masses between 100 eV and 100 keV. We find that, due to the suppression of the hydrogen atomic form factor at high energies, inelastic scattering provides the dominant constraint for DM masses above the keV scale. At lower masses, hydrogen ionization through DM absorption is the leading channel, driven by the higher efficiency of post-recombination energy injection in modifying the free-electron fraction. Although the bounds we derive are considerably weaker than existing laboratory and astrophysical limits, they provide a robust and independent cosmological probe of leptophilic DM interactions.
\end{abstract}

\maketitle

\tableofcontents


\section{Introduction}

The fundamental nature of dark matter (DM) remains one of the central open questions in modern cosmology and particle physics. While terrestrial experiments and astrophysical searches have placed increasingly stringent bounds on DM interactions with Standard Model (SM) particles, cosmological observations have emerged as a powerful and complementary probe of the particle properties of DM. The early Universe provides a pristine environment to study non-gravitational interactions between dark and visible matter, free from the uncertainties that affect the interpretation of laboratory and astrophysical searches.

The anisotropies of the cosmic microwave background (CMB) are an especially robust probe of DM–SM interactions, as they are well understood within linear cosmological perturbation theory and are very sensitive to modifications in the thermal and ionization history that such exotic interactions may induce.

In particular, DM processes that inject energy into the primordial plasma---such as decay or annihilation into SM particles---increase the free-electron fraction and heat the baryons, leaving characteristic imprints on the CMB power spectra. Such effects have been extensively studied for a broad range of decay and annihilation products~\cite{Adams:1998nr,Chen:2003gz, Padmanabhan:2005es, Galli:2009zc, Slatyer:2009yq, Kanzaki:2009hf, Hutsi:2011vx, 2011PhRvD..84b7302G, Giesen:2012rp, 2012PhRvD..85d3522F, Slatyer:2012yq,  Cline:2013fm, Weniger:2013hja, 
 Lopez-Honorez:2013cua, Diamanti:2013bia, Madhavacheril:2013cna,Galli:2013dna, Slatyer:2015jla,Poulin:2016nat,Slatyer:2016qyl,Liu:2016cnk, Poulin:2016anj, Cang:2020exa, Kawasaki:2021etm,Liu:2023nct, Capozzi:2023xie, Xu:2024vdn, Montefalcone:2025nmm}, with constraints extending down to DM masses of a few eV, at the hydrogen ionization threshold~\cite{Xu:2024vdn}.

 The leading cosmological signature depends on the properties of the DM candidate, including its spin and the structure of its coupling to SM particles.  In this work, we focus on sub-MeV vector and axial-vector (spin 1) DM coupled trilinearly to electrons. Such particles arise naturally in well-motivated extensions of the SM featuring hidden gauge sectors or direct leptophilic couplings (e.g., see Refs.~\cite{Essig:2013lka, Alexander:2016aln, Fabbrichesi:2020wbt, Caputo:2026pdw}). In these models, the full DM abundance can be generated through non-thermal mechanisms such as gravitational production~\cite{Chung:1998ua, Ahmed:2020fhc, Kolb:2023ydq} or field misalignment~\cite{Nelson:2011sf, Agrawal:2018vin}, 
independent of the DM-electron coupling studied here, ensuring the DM remains cold and is consistent with bounds on additional thermalized species at BBN~\cite{Boehm:2002yz,Serpico:2004nm,Nollett:2013pwa,Steigman:2014pfa,Nollett:2014lwa,Escudero:2018mvt,Sabti:2019mhn,Giovanetti:2021izc,An:2024nsw}.

For our models of interest, tree-level decay into an electron-positron pair is kinematically forbidden, while the loop-induced decay into two photons---which was the dominant energy injection channel for spin-0 leptophilic DM in our previous work~\cite{Montefalcone:2025nmm}---is forbidden by angular momentum conservation.  The dominant source of energy injection is instead a qualitatively distinct class of processes from those that are typically studied: DM conversion into SM radiation via interaction with a SM electron. Within our framework, the two relevant conversion channels are DM inelastic scattering with a free electron and DM absorption by bound atomic electrons, leading to hydrogen ionization, akin to the photoelectric effect.

Here, we study for the first time the imprint of these DM absorption processes on CMB anisotropies and derive constraints on the vector and axial-vector DM–electron coupling constants over the mass range $100\,\mathrm{eV} \leq m_{\rm DM} \leq 100\,\mathrm{keV}$,
using \textit{Planck} 2018 temperature, polarization, and lensing data~\cite{Planck:2019nip}. To self-consistently track the thermal and ionization evolution of the primordial plasma, we use a modified version of the \texttt{DarkHistory} code~\cite{Liu:2019bbm,Liu:2023fgu}, extended to include the DM conversion channels, interfaced with the cosmological Boltzmann solver \texttt{CLASS}~\cite{Blas:2011rf}.

 We obtain bounds that are considerably weaker than existing laboratory and astrophysical limits; nonetheless, our results provide the only constraints in the sub-MeV mass regime that rely exclusively on cosmological observables, offering a robust and independent probe of leptophilic DM interactions. We further note that, in the axial-vector case, limits in the literature exist only for DM masses above the MeV scale~\cite{Baruch:2022esd}; while recasting the relevant vector bounds into axial-vector form would likely also yield stronger constraints than ours, such a recasting has not been carried out in the literature and is beyond the scope of this work.
 
 The size of the couplings constrained here are, in the absence of additional dynamics, large enough to thermalize the DM with the SM plasma prior to BBN, which is inconsistent with our nonthermal assumption. A possible resolution is to invoke a phase transition in the dark sector that suppresses these rates at early times~\cite{Cohen:2008nb,Baker:2019ndr, Cohen:2008nb,Elor:2021swj, Das:2023enn, Mandal:2022yym} or, for sub-keV vector DM, an in-medium suppression that arises from kinetic mixing with the SM photon~\cite{Redondo:2008ec,Hardy:2016kme, An:2013yfc, Redondo:2013lna,Knapen:2017xzo}. Both mechanisms leave the late-time absorption processes constrained in this work unaffected.

This paper is organized as follows. 
In Sec.~\ref{sec:models}, we introduce the vector and axial-vector DM models under consideration and provide the corresponding DM conversion cross sections. 
In Sec.~\ref{sec:energy_injection}, we describe the methodology for calculating the effects of exotic energy injection on the CMB anisotropies, focusing on our use of \texttt{DarkHistory}, and outline our analysis framework in Sec.~\ref{sec:analysis}.
We present our constraints on DM-electron couplings using CMB temperature and polarization anisotropies in Sec.~\ref{sec:results_comparison}.
Finally, we conclude in Sec.~\ref{sec:conclusions}.
Throughout this work, we use natural units: $c=\hbar=1$.


\section{Models}
\label{sec:models}

We consider vector~$V$ and axial-vector~$A$ particles that couple to electrons and constitute the entirety of DM in the Universe.
As in our previous work~\cite{Montefalcone:2025nmm}, we focus on the sub-MeV mass regime with $m_{\rm DM}\lesssim m_e$, where~$m_e$ is the mass of the electron, ensuring that tree-level decay into an electron-positron pair is kinematically forbidden.
Unlike our previous scalar and pseudo-scalar DM scenarios~\cite{Montefalcone:2025nmm}, vector and axial-vector interactions do not permit a loop-induced decay into two photons because of angular momentum conservation. The leading decay channel
is instead the loop-induced decay into three photons, which is suppressed by~$\alpha^2_\mathrm{EM}$, where~$\alpha_\mathrm{EM}$ is the electromagnetic fine-structure constant, compared to the conversion processes of interest for this work.
Thus, we neglect any DM decay process.

The dominant low-energy signatures of these DM candidates are processes in which DM is converted into SM radiation through interactions with free or bound electrons. 
We remain agnostic about the mechanism responsible for setting the DM relic abundance, assuming only that it is generated through a process independent of the electron couplings studied here, without specifying the UV completion of the theory.
Additionally, we have verified that, at the level of the DM–electron couplings probed in this work, the conversion processes do not significantly alter the DM number density, with their rates staying below the Hubble expansion rate at temperatures $T\lesssim m_{\rm DM}$. The same coupling also permits elastic DM–electron scattering, but the corresponding rate is suppressed by an additional factor of $g_{(V/A)\bar ee}^2/\alpha_\text{EM}$ relative to the conversion processes considered here and is therefore negligible across the parameter space of interest.

We assume that DM remains thermally decoupled from the SM plasma at temperatures $T\gtrsim\,\mathrm{MeV}$, since thermalization would erase any pre-existing non-thermal abundance, produce a hot relic incompatible with structure formation, and contribute additional relativistic degrees of freedom in tension with bounds from light-element abundances~\cite{Boehm:2002yz,Serpico:2004nm,Nollett:2013pwa,Steigman:2014pfa,Nollett:2014lwa,Escudero:2018mvt,Sabti:2019mhn,Giovanetti:2021izc,An:2024nsw}.
In the sub-MeV mass regime, the relevant thermalization channels are $\gamma \,e^- \to e^-\,(V/A)$ at $T\gtrsim m_{\rm DM}$, and  
$e^+\,e^- \to \gamma\, (V/A)$ at $T \gtrsim m_e$, which become efficient for DM–electron coupling constants $\gtrsim 5\times 10^{-10}$, consistent with analogous results obtained for scalar DM~\cite{Knapen:2017xzo}.

As we show in Sec.~\ref{sec:results_comparison}, the upper bounds derived from our CMB analysis lie one to three orders of magnitude above this thermalization threshold; we defer the discussion of model-building mechanisms that reconcile our bounds with the non-thermal assumption to that section.


\subsection{Vector}

\begin{figure}[t]
  \centering
  \tikzset{every fermion/.style={very thick}}
  \begin{tikzpicture}
    \begin{feynman}
      \vertex (e_in_s)  at (-1, 1) {\(e\)};
      \vertex (X_in_s)  at (-1,-1) {$V/A$};
      \vertex[dot] (vs1) at (0, 0.0) {};
      \vertex[dot] (vs2) at (1.5, 0.0) {};
      \vertex (g_out_s) at (2.5, 1) {$\gamma$};
      \vertex (e_out_s) at (2.5,-1) {$e$};
      \diagram*{
        (e_in_s) -- [fermion] (vs1),
        (X_in_s) -- [scalar]   (vs1),
        (vs1)    -- [fermion] (vs2),   
        (vs2)    -- [photon]  (g_out_s),
        (vs2)    -- [fermion] (e_out_s),
      };
      \vertex (e_in_t)  at (4, 1.75) {$e$};
      \vertex (X_in_t)  at (4,-1.75) {$V/A$};
      \vertex[dot] (vt1) at (5, 0.75 ) {};
      \vertex[dot] (vt2) at (5,-.75) {};
      \vertex (g_out_t) at (6, 1.75) {\(\gamma\)};
      \vertex (e_out_t) at (6,-1.75) {\(e\)};

      \diagram*{
        (e_in_t) -- [fermion] (vt1),
        (vt1)    -- [photon]  (g_out_t),
        (vt1)    -- [fermion] (vt2),   
        (X_in_t) -- [scalar]   (vt2),
        (vt2)    -- [fermion] (e_out_t),
      };
    \end{feynman}
  \end{tikzpicture}
  \caption{Feynman diagrams for tree-level DM-electron inelastic scattering, in which a vector~($V$) or axial-vector~($A$) DM particle is converted into a photon.}
  \label{fig:feynman_diagrams}
\end{figure}
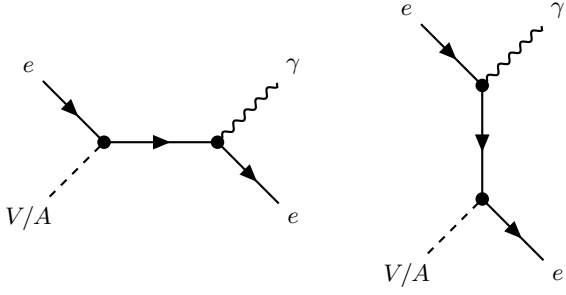

The interaction Lagrangian for a vector DM particle~$V^\mu$ coupled to electrons is
\begin{equation}
    \mathcal{L}_{\rm int} = g_{V \bar e e} V^\mu \bar \psi_e  \gamma_\mu \psi_e\,, \label{eq:Lint_Vscat}
\end{equation}
where~$g_{V\bar e e}$ is the vector DM-electron coupling constant. 
Inelastic scattering with an electron induces the conversion of $V^\mu$ into a photon, as illustrated in Fig.~\ref{fig:feynman_diagrams}, with corresponding 
cross section
\begin{align}
    \sigma_{Ve^{-}\rightarrow \gamma e^{-}} &\approx \frac{\alpha_{\rm EM}\,g_{V\bar e e}^2}{3\,v_r\,m_e^2}\left(\frac{3\,m_{\rm V}+2\,m_e}{3\,m_{\rm V}+m_e} \right)+\mathcal{O}(v^2_r) ,
    \label{eq:sV_scat}
\end{align}
where~$v_r$ is the DM-electron relative velocity, and we assume $m_{\rm V}\ll m_e$ (see Appendix~\ref{app:DM_crosssection} for the full derivation).

After recombination and before reionization, the vast majority of electrons reside in neutral atoms.
A bound atomic electron can absorb the vector DM particle, resulting in the ionization of hydrogen and ejection of the electron into the continuum. The ionization cross section is
\begin{equation}
    \sigma_{V+H\rightarrow e^{-}+p^{+}}\approx g^2_{V\bar e e}Q_{V}(\epsilon)a^2_0/v_r, 
    \label{eq:sV_Hion}
\end{equation}
where $a_0\approx 5.29\times 10^{-11}\,{\rm cm}$ is the Bohr radius and $\epsilon$ is the energy of the incoming DM particle. 
The atomic form factor
\begin{equation}
    Q_V(\epsilon) = \frac{\pi}{\epsilon\,a_0^2}\sum_{bc}\left|\int e^{i\vec{k} \cdot \vec{r}} \left(\psi_{e,b}^\dagger\,\mathbb{I}\,\psi_{e,c} \right)\,d^3r\right|^2
    \label{eq:QV}
\end{equation}
describes the overlap between the initial bound-state electron wavefunction $\psi_{e,b}$ and the final continuum state $\psi_{e,c}$, thereby capturing the energy dependence of the ionization process. The sum runs over all initial bound states~$b$ and final continuum states~$c$, with $\mathbb{I}$ denoting the identity matrix in spinor space  and~$|\vec{k}|=\sqrt{\epsilon^2-m_V^2}$ is the momentum of the incoming DM particle. 
We provide additional details on the derivation and evaluation of~$Q_V(\epsilon)$ in Appendix~\ref{app:hydrogen_ionization_formfactor}.

\begin{figure}[t]
    \centerline{
    \includegraphics[width=\linewidth]{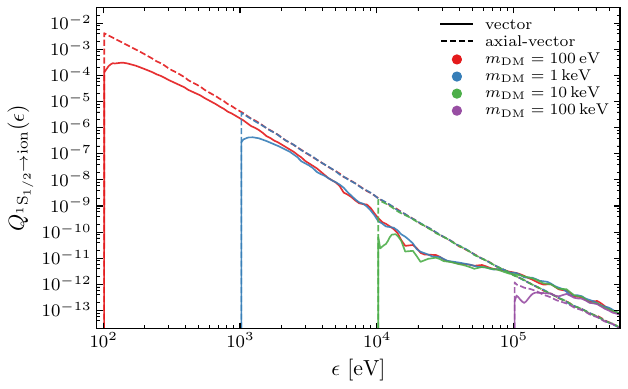}}
    \caption{Hydrogen atomic form factor $Q$ for ionization from its ground state level ($^1S_{1/2}$) as a function of the incoming DM energy  $\epsilon$, computed for
$m_{\rm DM}=\{100\,{\rm eV},\,1\,{\rm keV},\,10\,{\rm keV},\,100\,{\rm keV}\}$, shown in the colors indicated in the legend. Solid curves correspond to the vector interaction, while dashed curves show the axial-vector case. In all scenarios, the form factor peaks at $\epsilon \approx m_{\rm DM}$ and is increasingly suppressed at larger energies and DM masses, reflecting the mismatch between the energy transfer and the $13.6\,{\rm eV}$ hydrogen binding energy.}
\label{fig:Q_form_factor}
\end{figure}

In Fig.~\ref{fig:Q_form_factor}, we show the ionization form factor for $m_V=\{100\,{\rm eV},\,1\,{\rm keV},\,10\,{\rm keV},\,100\,{\rm keV}\}$, spanning the DM mass range we consider in this work. $Q_V(\epsilon)$ peaks at the kinematic threshold $\epsilon \approx m_V$ and falls rapidly at higher $\epsilon$, with its overall amplitude becoming increasingly suppressed as the DM mass grows. This behavior reflects the fact that hydrogen ionization is most efficient when the energy transferred to the bound electron is comparable to its binding energy of $13.6\,$eV for the $^1S_{1/2}$ ground state. For the DM mass range considered here, $m_V\gg 13.6\,{\rm eV}$ and the form factor therefore lies entirely in the suppressed tail of the ionization response, with the suppression deepening for larger $m_V$. This behavior is qualitatively analogous to that of standard hydrogen photoionization, where the cross section peaks at the ionization threshold and falls as $\sigma_{\rm photo}\propto E_\gamma^{-7/2}$ above it~\cite{bethe:photoionization,sobelman:photoionization}, where $E_\gamma$ is the energy of the ionizing photon.

Moreover, in Fig.~\ref{fig:Q_form_factor}, we report only the contribution from ionization of the~$^1S_{1/2}$ bound state,  which dominates the form factor in the DM mass range of interest, with contributions from excited bound states strongly suppressed in this regime (see Fig.~\ref{fig:A2} in Appendix~\ref{app:hydrogen_ionization_formfactor}). This hierarchy between bound states qualitatively analogous to traditional photoionization~\cite{bethe:photoionization, sobelman:photoionization}. Moreover, outside the narrow window of recombination, essentially all hydrogen atoms reside in the ground state; thus, we retain only the $^1S_{1/2}$ contribution in our calculations.


\subsection{Axial-vector}

The interaction Lagrangian for an axial-vector DM particle~$A^\mu$ coupled to electrons is
\begin{equation}
    \mathcal{L}_A = g_{A \bar e e} A^\mu \bar \psi_e  \gamma_5 \gamma_\mu \psi_e\,,
\end{equation}
where~$g_{A\bar e e}$ is the axial-vector DM-electron coupling constant.
The dominant energy injection channels in the sub-MeV mass range are the same as the vector case: DM conversion due to inelastic scattering with free electrons and absorption by neutral hydrogen.
The inelastic cross section for axial-vector DM scattering with an electron is
\begin{equation}
    \sigma_{A\,e^{-}\rightarrow\gamma\,e^{-}} \approx \frac{4\,\alpha_{\rm EM}\,g_{A\bar e e}^2}{3\,v_r\,m_e^2}\left(\frac{m_{\rm A}+\,m_e}{3\,m_{\rm A}+m_e} \right)+\mathcal{O}(v^2_r) , 
    \label{eq:sA_scat}
\end{equation}
where, as in the vector case, we have expanded the result in the relative particle velocity $v_r$ and assumed $m_A\ll m_e$. The complete derivation is reported in Appendix~\ref{app:DM_crosssection}.

Similarly, the hydrogen ionization cross section can be expressed in an
analogous form to the vector case:
\begin{equation}
    \sigma_{A+H\rightarrow e^{-}+p^{+}} \approx g^2_{A\bar e e} Q_{A}(\epsilon) a^2_0/v_r,
    \label{eq:sA_Hion}
\end{equation}
where the corresponding atomic form factor~$Q_A(\epsilon)$ differs from that of the vector case only because the axial-vector structure of the electron current:
\begin{equation}
    Q_A(\epsilon) = \frac{\pi}{\epsilon\,a_0^2}\sum_{bc} \left|\int e^{i\mathbf{k} \cdot \mathbf{r}} \left(\psi_{e,b}^\dagger\,\gamma_5\,\psi_{e,c} \right)\,d^3r\right|^2. 
    \label{eq:QA}
\end{equation}
The form factor is shown as dashed lines in Fig.~\ref{fig:Q_form_factor}. We find the same qualitative behavior as the vector case: the form factor $Q_A(\epsilon)$ peaks at the threshold energy $\epsilon\approx m_A$ and is strongly suppressed as the DM mass increases. Relative to the vector scenario, the axial-vector form factor is more sharply peaked and exhibits less numerical noise. This difference originates from the structure of the DM–electron interaction. For axial-vector couplings, the leading contribution in a multipole expansion of $Q_A(\epsilon)$ is the monopole term, while in the vector case, the dominant contribution arises from the dipole term, as discussed in greater detail in Appendix~\ref{app:hydrogen_ionization_formfactor}. As in the vector case, we report only the contribution from the bound
state $^1S_{1/2}$, since contributions from excited states remain subdominant in the DM mass range relevant for our analysis.


\section{Exotic Energy Injection}
\label{sec:energy_injection}


\subsection{Energy injection and deposition}

The rate of energy injection into the primordial plasma
from DM conversion into a photon through inelastic
scattering is
\begin{equation}
    \left(\frac{d^2E}{dVdt}\right)_{{\rm inj}} = E_{\rm inj}\,n_e\,n_{\rm DM}\langle \sigma v\rangle_{(V/A)\, e^{-}\rightarrow \gamma e^{-}},
\end{equation}
where~$E_{\rm inj}$ is the total energy injected, and~$n_e$ and~$n_{\rm DM}$ are the number densities of free electrons and DM, respectively. 
The thermally averaged cross section $\langle \sigma v\rangle_{(V/A)\, e^{-}\rightarrow \gamma e^{-}}$ in the
non-relativistic limit, relevant at late times, is approximately
\begin{equation}
    \langle \sigma v\rangle_{(V/A)\, e^{-}\rightarrow \gamma e^{-}}\approx \sigma_{(V/A)\, e^{-}\rightarrow \gamma e^{-}}v_r,
\end{equation}
where~$\sigma_{(V/A)\,e^{-}\rightarrow \gamma\,e^{-}}$ is given in Eqs.~\eqref{eq:sV_scat} and~\eqref{eq:sA_scat} for vector and axial-vector DM, respectively. 

Each scattering event injects both a photon and an electron into the plasma. 
In the DM mass range of interest, most of the injected energy is carried away by the photon, but the electron also receives a non-negligible momentum kick which, particularly at large DM masses, can push it outside its thermal distribution, significantly affecting the subsequent energy deposition process.\footnote{
A naive estimate suggests that at recombination the ratio of the momentum, $p_e$, of an electron that has undergone inelastic scattering to its typical thermal momentum scale, $\sqrt{m_e T_b}$, is given by $\sim 10^3\, m_\mathrm{DM}/m_e$, where we take $p_e\approx m_{\rm DM}$ from Eq.~\eqref{eq:Einj_scat} and $T_b\approx 10^{-6}\,m_e$ at recombination. This ratio increases at lower redshifts as $T_b$ decreases and is generically greater than unity for a large portion of the DM mass range of interest, confirming that the scattered electron carries a substantial amount of kinetic energy  whose deposition into the plasma must be carefully accounted for.} We account for this by treating the photon and the kicked electron as separate injection products, with energies fixed by momentum conservation (see Appendix~\ref{app:DM_in_DarkHistory}).
The total injected energy is
\begin{equation}
    E_{\rm inj} = p_\gamma + K_e\approx p_\gamma + \frac{p_\gamma^2}{2m_e}, \label{eq:Einj_scat}
\end{equation}
where $K_e$ is the kinetic energy of the outgoing free-electron and~$p_\gamma\approx m_{\rm DM}(m_{\rm DM}+2m_e)/[2(m_{\rm DM}+m_e)]$.
The complete derivation of these results is provided in Appendix~\ref{app:DM_in_DarkHistory}.

For DM absorption associated with hydrogen ionization, the energy injection rate per unit volume is
\begin{equation}
    \left(\frac{d^2E}{dVdt}\right)_{{\rm inj}} = n_{\rm HI}\,\rho_{\rm DM}\,\langle \sigma v\rangle_{(V/A)\,H \rightarrow p^{+}\,e^{-}},
\end{equation}
where~$n_{\rm HI}$ is the number density of neutral hydrogen atoms, and we approximate the injected energy~$E_{\rm inj} \approx m_{\rm DM}$ as being entirely carried away by the outgoing electron in the form of kinetic energy. 
This treatment neglects the contribution of the recoiling proton, as it receives a negligible fraction of the injected energy in the DM mass range of interest, and protons deposit energy into the plasma less efficiently than electrons~\cite{Liu:2019bbm}.

The thermally averaged cross section $ \langle \sigma v\rangle_{(V/A)\,H\rightarrow p^{+}\,e^{-}}$ in the non-relativistic limit is
\begin{equation}
    \langle \sigma v\rangle_{(V/A)\,H\rightarrow p^{+}\,e^{-}} \approx g_{(V/A)\,\bar e e}^2 a_0^2 Q_{V/A}(m_{\rm DM}),
\end{equation}
 where $Q_{(V/A)}(m_{\rm DM})$ is the atomic ionization form factor introduced in the previous section, Eqs.~\eqref{eq:QV} and~\eqref{eq:QA}, evaluated at energy $\epsilon = m_{\rm DM}$.
See Appendix~\ref{app:DM_absorption_rate} for the complete derivation.

Having specified the energy injection rates associated with the DM conversion processes discussed above, we now turn to the subsequent deposition of this energy into the primordial plasma. Once injected, the energy carried by photons and electrons is redistributed among several physical channels, including hydrogen and helium ionization, Lyman-$\alpha$ excitation, heating of the intergalactic medium, and the production of low-energy continuum photons. The physics of these radiation transfer processes is conveniently encapsulated by the energy deposition functions $f_c(z)$, which describe the fraction of energy deposited into each channel $c$:
 \begin{equation}
     \left(\frac{d^2E}{dV\,dt}\right)_{{\rm dep},\,c}
= f_c(z)\left(\frac{d^2E}{dV\,dt}\right)_{\rm inj}.
 \end{equation}

 We compute the energy deposition functions $f_c(z)$ using a modified version of \texttt{DarkHistory v2.0}~\cite{Liu:2023fgu}, which we modify to include the conversion processes relevant for the vector and axial-vector DM models of interest. A summary of the main modifications to the \texttt{DarkHistory} code is provided in Appendix~\ref{app:DM_in_DarkHistory}, together with representative visualizations of $f_c(z)$.

 
\subsection{Effects on the ionization history and CMB anisotropies}

\begin{figure}[t]
    \centerline{
    \includegraphics[width=\linewidth]{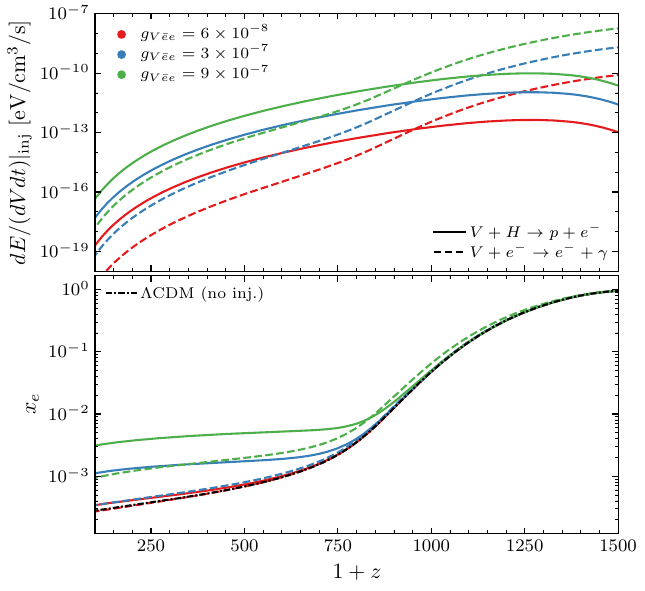}}
    \caption{{\it Top panel}: Energy density injection rates as a function of redshift for a representative case of vector DM conversion with $m_{\rm DM}=1\,$keV. Solid lines correspond to DM absorption by neutral hydrogen, while dashed lines correspond to DM inelastic scattering with free electrons.
    Results are shown for three benchmark values of the DM–electron coupling~$g_{V\bar e e}$, corresponding to the {\it Planck} sensitivity limit derived in this work  and to couplings five and fifteen times larger, shown in \textcolor{tabred}{red}, \textcolor{tabblue}{blue}, and \textcolor{tabgreen}{green}, respectively.
    {\it Bottom panel}: Corresponding evolution of the ionization fraction, $x_e(z)$, for the same cases shown in the {\it Top panel}. The fiducial $\Lambda$CDM ionization history in the absence of any energy injection is shown as the black dot-dashed line.
    The markedly different redshift dependence of the two energy injection processes leads to distinct impacts on the ionization history. DM absorption by hydrogen enhances $x_e$ for $z\lesssim 500$, while inelastic scattering produces a slight enhancement already at high redshifts and a more pronounced effect around recombination.}
    \label{fig:energyinjection_xe}
\end{figure}

To quantify the impact of these exotic energy injection mechanisms on CMB anisotropies, we follow Refs.~\cite{Liu:2023nct, Capozzi:2023xie, Xu:2024vdn, Montefalcone:2025nmm} and supply the energy deposition functions $f_c(z)$ computed with \texttt{DarkHistory} to the \texttt{injection} module of \texttt{CLASS}, which self-consistently evolves the ionization fraction and baryon temperature in the presence of additional energy deposition~\cite{Stocker:2018avm}.

The primary cosmological effect of the DM conversion processes considered here is the enhancement of the free-electron fraction over a broad range of redshifts after recombination, relative to $\Lambda$CDM. 
We illustrate this behavior in Fig.~\ref{fig:energyinjection_xe}, which displays the redshift evolution of the energy injection rates and their corresponding effects on the ionization history for a vector DM particle with $m_{\rm DM}=1\,$keV and three representative values of the DM-electron coupling constant~$g_{V\bar e e}$. Specifically, we choose $g_{V\bar e e}$~values corresponding to the {\it Planck} sensitivity limit derived in this work and to couplings five and fifteen times larger, to make the impact on the ionization history more readily visible.

In the case of DM absorption by hydrogen,
shown as the solid lines in Fig.~\ref{fig:energyinjection_xe}, the free-electron fraction is enhanced relative to $\Lambda$CDM only at relatively low redshifts, $z\lesssim 500$, with the enhancement becoming larger in amplitude and shifting to higher redshifts as the DM–electron coupling increases.
This behavior follows from the redshift dependence of both the energy injection and deposition mechanism and the ionization efficiency of the plasma. 
The main impact occurs post recombination, when the energy injected acts on a Universe that is already largely neutral, making it particularly efficient at producing additional free electrons that persist over extended periods.  This behavior is especially relevant for
absorption through hydrogen ionization, which has no
effect prior to recombination, when neutral hydrogen is
effectively absent.

On the other hand, DM conversion into a photon via inelastic scattering with a free electron, shown with dashed lines in Fig.~\ref{fig:energyinjection_xe}, already increases the global free-electron fraction well before recombination ends. 
As a result, there is a visibly different redshift dependence of the energy injection rate, relative to
DM absorption. In fact, the inelastic scattering rate becomes strongly suppressed after recombination, as it is proportional to the free-electron fraction, which drops sharply at recombination. Since the impact of inelastic scattering is significantly reduced at low redshifts,  its effect on the ionization history is concentrated around recombination, even though, at that epoch, the rapid recombination rate makes injected energy less efficient at producing long-lived free electrons than at later times.

\begin{figure*}[ht]
    \centerline{
    \includegraphics[width=\linewidth]{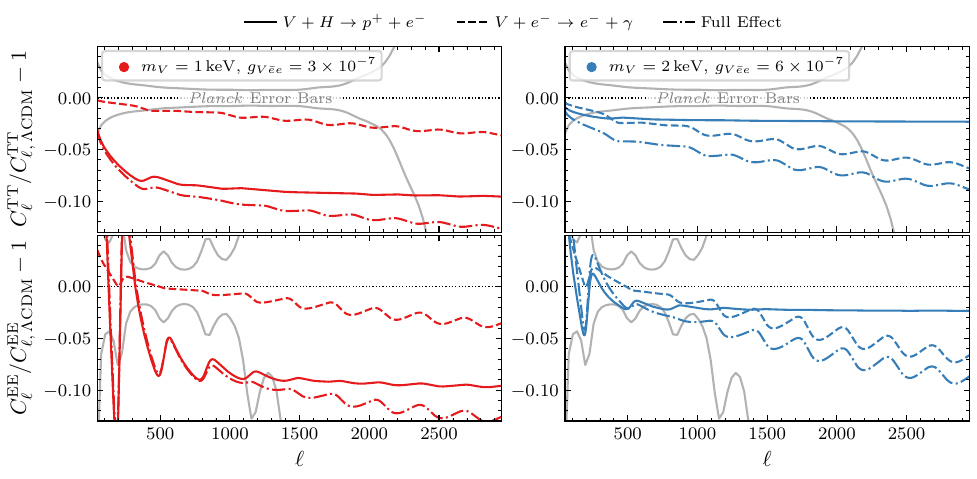}}
    \caption{CMB temperature ({\it top panels}) and polarization ({\it bottom panels}) power spectrum residuals for energy injection from vector DM, relative to a fiducial $\Lambda$CDM cosmology with no exotic energy injection. 
    Solid lines show the contribution from DM absorption leading to hydrogen ionization, dashed lines correspond to DM inelastic scattering with a free electron, and dot-dashed lines indicate the combined effect of these two processes. 
    Finally, for visual comparison, we display in \textcolor{tabgray}{gray} the binned {\it Planck} measurement uncertainties for the corresponding spectra at each multipole.
    The distinct redshift dependence of the two energy injection processes leads to qualitatively different imprints in the CMB spectra. Both produce an overall suppression due to an increased optical depth, and the inelastic scattering scenario additionally affects the acoustic peak locations and the diffusion damping scale. 
    While the relative importance of the two injection processes depends on the DM mass, in all cases the combined effect is well described by the superposition of the two individual signals.
    We set $m_{\rm DM}=1\,{\rm keV}$ and $g_{V\bar e e}=3\times 10^{-7}$ ({\it left panels}) and $m_{\rm DM}=2\,$keV and $g_{V\bar e e}=6\times 10^{-7}$ ({\it right panels}), where the chosen couplings correspond approximately to five times the {\it Planck} sensitivity limit we derive in this work.}
    \label{fig:effect_on_CMB}
\end{figure*}

Modifications of the ionization history leave characteristic imprints on the temperature and polarization anisotropies of the CMB. We illustrate these effects in Fig.~\ref{fig:effect_on_CMB}, which shows the CMB spectra residuals relative to a fiducial $\Lambda$CDM cosmology for two representative vector DM scenarios with $m_{\rm DM}$ equal to $1\,$keV and $2\,$keV in the left and right panels, respectively.
The DM-electron coupling is set to approximately five times larger than the {\it Planck} sensitivity limit derived in this work to better emphasize the features created by these exotic energy injections. 

For DM absorption through hydrogen ionization, the dominant effect is an overall suppression of the CMB spectra, due to the increase in the optical depth over a large redshift range. For DM inelastic scattering, there is also a constant suppression associated with the increased optical depth. Additionally, there is a non-negligible impact from the enhanced free-electron fraction during recombination, which leads to a broadening of the surface of last scattering and a slight delay of recombination. As a consequence, as clearly illustrated in Fig.~\ref{fig:effect_on_CMB}, the acoustic peak locations are shifted due to the minor increment in the baryon sound speed during this epoch, while the delayed decoupling enhances photon diffusion and results in stronger Silk damping at small angular scales.

Finally, in the same figure, we also display as dot-dashed lines the CMB power spectra residuals resulting from the combined effect of both energy injection processes. The total signal roughly follows a linear superposition of the individual contributions, consistent with the weak-interaction regime~\cite{Ali_Ha_moud_2024} in which the cross section is sufficiently small that each process contributes linearly and independently to the deposited energy sourcing the evolution of the ionization fraction.


\section{Analysis}
\label{sec:analysis}

We derive upper bounds on the DM–electron coupling constant by performing three separate MCMC analyses corresponding to each individual energy injection channel, i.e.  DM absorption and DM inelastic scattering, as well as their combined effect. 
In all cases, we place constraints on the vector and axial-vector DM–electron coupling over the DM mass range $100\,{\rm eV}\leq m_{\rm DM}\leq 100\,{\rm keV}$ using \textit{Planck} 2018 temperature, polarization, and lensing data.

We employ the~\textit{Planck} 2018 PR3 likelihood code, using the \texttt{plik-lite} likelihood for high multipoles ($\ell \geq 30$), the~\texttt{commander} and~\texttt{SimAll} likelihoods at low multipoles ($2 \leq \ell \leq 29$) for the temperature and polarization power spectra, respectively, and the~\texttt{SMICA} lensing reconstruction likelihood~\cite{Planck:2019nip}. 
The energy deposition functions are precomputed by~\texttt{DarkHistory} for discrete values of the DM mass and coupling under the assumption of a fixed fiducial cosmology, and we interpolate to obtain intermediate values of the coupling during MCMC sampling. 
Although these functions are in principle mildly sensitive to cosmological parameters, we explicitly verified in Ref.~\cite{Montefalcone:2025nmm} that altering the fiducial cosmology in \texttt{DarkHistory} during MCMC sampling leads to equivalent constraints on the energy injection. 

The MCMC analyses are performed using the Boltzmann solver \texttt{CLASS v3.2.5} interfaced with the sampler \texttt{MontePython}~\cite{Audren:2012wb, Brinckmann:2018cvx}.
The angular size of the sound horizon at decoupling~$\theta_s$, the physical baryon density~$\omega_b$, the physical matter density~$\omega_m$, the scalar amplitude~$\ln(10^{10}A_s)$, the scalar spectral index~$n_s$, and the redshift at reionization~$z_{\rm{reio}}$ are all sampled with broad uniform priors.
Additionally, for the fixed values of DM masses $m_\mathrm{DM}=\{100\,\rm{eV},\, 1\,\rm{keV},\, 10\,\rm{keV},\, 100\,\rm{keV}\}$, we sample $\log_{10} g_{V \bar{e}e}$ and/or~$\log_{10} g_{A \bar{e}e}$ with a uniform prior. 
Convergence is established using the Gelman-Rubin criterion~\cite{Gelman:1992zz} requiring that~$R < 0.01$, and the resulting chains are analyzed using \texttt{GetDist}~\cite{2019arXiv191013970L}.

For the individual absorption channels, we run MCMC analyses only for the vector case and obtain the corresponding axial-vector constraints by rescaling, exploiting the simple mapping between the two scenarios at the level of the cross sections, Eqs.~\eqref{eq:sV_scat} and~\eqref{eq:sA_scat} for inelastic scattering and Eqs.~\eqref{eq:sV_Hion} and~\eqref{eq:sA_Hion} for hydrogen ionization. The combined-channel constraints, in contrast, are derived from independent MCMC runs for the vector and axial-vector models, since the two cases exhibit a mildly different DM mass dependence in their respective cross sections, which shifts the relative importance of the hydrogen ionization and inelastic scattering channels and thus affects the total combined signal.


\section{Results and Comparison with Other Constraints}
\label{sec:results_comparison}

\begin{figure*}[ht]
    \centerline{
    \includegraphics[width=1\linewidth]{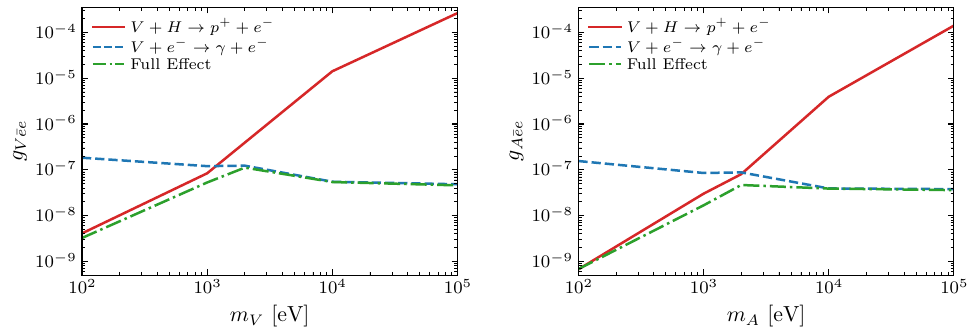}}
    \caption{\textit{Planck} 95\% C.L. upper limits on the dimensionless DM-electron coupling for the vector ({\it left panel}) and axial-vector ({\it right panel}) cases as a function of the DM mass. 
    In each panel, the \textcolor{tabred}{red} solid curve shows the constraints derived from DM absorption through hydrogen ionization, the \textcolor{tabblue}{blue} dashed curve corresponds to DM conversion into a photon via inelastic scattering, and the \textcolor{tabgreen}{green} dot–dashed curve shows the combined effect of the two energy injection processes. 
    Inelastic scattering provides the dominant constraint at all masses for axial-vector DM and at $m_{\rm DM}\gtrsim 1\,$keV for vector DM, while DM absorption through hydrogen ionization becomes relevant only at low vector masses, enhancing the combined constraining power of the two process.}
    \label{fig:upperlimits_on_gVAee}
\end{figure*}

We present the~$95\%$ confidence level (C.L.) upper limits on the dimensionless DM-electron coupling constant as a function of the DM mass in Fig.~\ref{fig:upperlimits_on_gVAee}. 
Vector and axial-vector DM upper limits are reported in the left and right panels of the figure, respectively. The solid \textcolor{tabred}{red} curves show the constraints obtained from DM absorption through hydrogen ionization, the dashed \textcolor{tabblue}{blue} curves correspond to DM conversion into a photon via inelastic scattering, and the dot–dashed \textcolor{tabgreen}{green} lines represent the combined constraint of these two energy injection processes.

For DM masses~$m_{\rm DM}\gtrsim 1\,$keV, the dominant source of constraints arises from DM inelastic scattering. The contribution from hydrogen ionization in this regime is negligible due to the strong suppression of the corresponding
atomic form factor at such high energies (see~Fig.~\ref{fig:Q_form_factor}). The resulting constraints from inelastic scattering are approximately constant as a function of the DM mass, reflecting the fact that the corresponding cross sections are largely independent of $m_{\rm DM}$, cf.~Eqs.~\eqref{eq:sV_scat} and~\eqref{eq:sA_scat}. A mild weakening of the constraints is observed toward lower DM masses, which can be attributed to the reduced efficiency of sub-keV energy injection~\cite{Liu:2019bbm,Liu:2023fgu}.

At lower DM masses,~$m_{\rm DM}\lesssim 1\,$keV, the contribution of the hydrogen ionization channel
quickly becomes the dominant source of the constraints, reflecting the reduced suppression of the hydrogen ionization form factor at these energies~(see Fig.~\ref{fig:Q_form_factor}), combined with the higher efficiency of post-recombination energy injection in modifying the free-electron fraction, which is when this channel is predominantly active.

In the intermediate regime, $m_{\rm DM}\sim 1\,$keV, the two individual processes yield comparable constraints, and it is thus important to account for the combined effect of both injection channels, which yields a mildly stronger constraint than that obtained from either injection process alone. To better capture the transition between the low- and high-mass regimes, we perform an additional analysis at $m_{\rm DM}=2\,$keV, which confirms the smooth interpolation between $1\,$keV and $10\,$keV.

While the vector and axial-vector models share the same qualitative features described above, they differ in quantitative detail. For sub-keV DM masses, the axial-vector constraints are slightly stronger than the vector constraints, due to the substantial enhancement in the axial-vector atomic form factor, as shown in Fig.~\ref{fig:Q_form_factor}. At large masses, where DM inelastic scattering dominates in both models, the resulting constraints are effectively equivalent. The transition between these two regimes occurs at $m_{\rm DM}\simeq 1\,$keV for vector DM and $m_{\rm DM}\simeq 2\,$keV for axial-vector DM, and the difference reflects the mildly shallower dependence of the axial-vector form factor on the DM mass.

\begin{figure}[t]
    \centerline{
    \includegraphics[width=\linewidth]{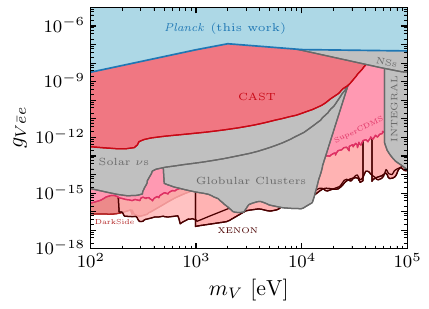}}
    \caption{Comparison of our {\it Planck} 95\% C.L. CMB upper limits on the vector DM-electron coupling $g_{V\bar e e}$ (in \textcolor{tabblue}{blue}) with limits from previous studies. 
    We show (in \textcolor{tabgray}{gray}) stellar cooling limits from neutron stars and globular clusters~\cite{Hong:2020bxo, Dolan:2023cjs}, constraints from INTEGRAL observations~\cite{Laha:2020ivk, Ferreira:2022egk}, and the solar luminosity/neutrino bound~\cite{Croon:2020lrf}.
    We also show in different shades of \textcolor{tabred}{red} the direct detection limits from DarkSide~\cite{DarkSide:2022knj}, XENON~\cite{XENON:2019gfn, XENON:2020rca, XENON:2021myl, XENONCollaboration:2022kmb}, and SuperCDMS~\cite{Aralis:2019nfa}, along with with helioscope constraints from CAST~\cite{Redondo:2008aa}.}
    \label{fig:Vector}
\end{figure}

In Fig.~\ref{fig:Vector}, we compare the upper limits on the vector DM–electron coupling, derived from our analysis that includes both energy injection channels of interest, with existing constraints in the literature.
We show in different shades of \textcolor{tabred}{red} the direct detection bounds from DarkSide~\cite{DarkSide:2022knj}, XENON~\cite{XENON:2019gfn, XENON:2020rca, XENON:2021myl, XENONCollaboration:2022kmb}, SuperCDMS~\cite{Aralis:2019nfa}, as well as helioscope constraints from CAST~\cite{Redondo:2008aa}, which probe vector particles produced in the Sun. 
Astrophysical bounds are shown in \textcolor{tabgray}{gray}, including constraints from INTEGRAL observations~\cite{Laha:2020ivk, Ferreira:2022egk}, stellar cooling limits from neutron stars~\cite{Hong:2020bxo} and globular clusters~\cite{Dolan:2023cjs}, as well as the solar luminosity/neutrino bound~\cite{Croon:2020lrf}.

Our constraints derived from CMB anisotropies, shown in \textcolor{tabblue}{blue}, are considerably weaker than existing laboratory and astrophysical bounds, by approximately eight to nine orders of magnitude compared to the strongest limits from the XENON collaboration. 

We do not present a corresponding comparison plot for the axial-vector case, since constraints in the literature exist only for DM masses above the MeV scale~\cite{Baruch:2022esd}. 
We expect that any of the vector DM bounds displayed in Fig.~\ref{fig:Vector} that can be consistently recast into axial-vector limits would also be more stringent than the CMB constraints derived here.

Finally, we revisit our assumption that DM, which we assume is produced non-thermally, remains non-thermal  throughout cosmic history. The couplings constrained in this work lie one to three orders of magnitude above the threshold for DM thermalization with the SM plasma prior to BBN, so consistency requires a mechanism that suppresses the thermalization rate at early times while leaving the late-time absorption processes unaffected. A simple and well-motivated resolution is to invoke a dark-sector phase transition that abruptly reduces the DM mass to its sub-MeV value, occurring after BBN but well before recombination~\cite{Cohen:2008nb,Baker:2019ndr,Elor:2021swj,Das:2023enn,Mandal:2022yym}. An additional possibility, specific to sub-keV vector DM, is to assume that the DM–electron coupling originates from kinetic mixing with the SM photon, in which case in-medium effects in the early-Universe plasma naturally suppress the effective coupling at temperatures $T\gg m_V$~\cite{Redondo:2008ec,Hardy:2016kme,An:2013yfc,Redondo:2013lna,Knapen:2017xzo}.

\section{Conclusions}
\label{sec:conclusions}

In this work, we derive constraints from CMB anisotropies on sub-MeV vector and axial-vector DM that couples leptophilically to the SM. These models give rise to two distinct channels of energy injection into the SM plasma at late times, namely DM conversion into a photon via inelastic scattering with free electrons and via hydrogen ionization from DM absorption. We perform separate MCMC analyses for each channel, as well as a combined analysis, using \textit{Planck} 2018 temperature, polarization, and lensing data to place 95\% C.L. upper limits on the DM–electron coupling constants $g_{V\bar ee}$
and $g_{A\bar{e}e}$ over the mass range $100\,\mathrm{eV}\leq m_{\rm DM}\leq 100\,\mathrm{keV}$.

We find that for DM masses $m_{\rm DM}\gtrsim 1\,$keV, the dominant constraint arises from DM inelastic scattering and is approximately mass-independent. 
At lower masses, the hydrogen ionization channel becomes the leading source of constraints, reflecting the enhanced atomic form factor and the higher efficiency of post-recombination energy injection. In the intermediate regime around $m_{\rm DM}\sim 1\,$keV, the two channels yield comparable individual limits and their combination produces a mildly stronger bound. While the vector and axial-vector models share these qualitative features, the axial-vector constraints are mildly stronger at sub-keV DM masses due to the enhancement of the corresponding atomic form factor, as shown in Fig~\ref{fig:Q_form_factor}.

Comparing our CMB bounds with existing laboratory and astrophysical constraints, we find that the limits derived here are considerably weaker. Nevertheless, the constraints presented in this work provide the only limits in this mass range that rely exclusively on cosmological observables, offering a probe of leptophilic DM interactions that is independent of the assumptions entering laboratory searches or stellar modeling.

At the same time, the couplings constrained in this work lie one to three orders of magnitude above the threshold for thermalization with the SM plasma prior to BBN, implying that a minimal scenario with unmodified DM properties throughout cosmic history is not self-consistent. Consistency with BBN can be naturally restored through well-motivated extensions such as a dark-sector phase transition at late times~\cite{Cohen:2008nb,Elor:2021swj,Baker:2019ndr} or, for sub-keV vector DM, a kinetic-mixing origin of the DM–electron coupling~\cite{Knapen:2017xzo}.

As a whole, our findings demonstrate that CMB anisotropies remain a sensitive probe of leptophilic DM interactions, even in scenarios that invoke nontrivial dark sector dynamics to keep DM cold and consistent with BBN constraints on additional thermalized species, providing an independent source of bounds on the DM–electron coupling through energy injection processes active around and beyond recombination. 

Together with our constraints on scalar and pseudo-scalar DM derived in Ref.~\cite{Montefalcone:2025nmm}, where the dominant energy injection mechanism is the loop-induced two-photon decay channel, our work shows that CMB anisotropies provide  a meaningful probe of leptophilic DM for both spin-0 and spin-1 candidates in the sub-MeV regime, with the spin-0 case yielding particularly competitive constraints. Notably, for the vector and axial-vector models studied here, this constraining power persists despite the absence of any loop-induced decay to photons, with the leading signatures arising entirely from the DM conversion processes analyzed in this work.


\begin{acknowledgments}
We thank Gilly Elor for early collaboration on this project. We are grateful to Yacine Ali-Ha\"{i}moud for useful discussions and to Masahiro Ibe, Wakutaka Nakano, and Yutaro Shoji for providing extensive support with the \texttt{cFAC} code.
We acknowledge the use of \texttt{CLASS}~\cite{Blas:2011rf}, \texttt{GetDist}~\cite{2019arXiv191013970L}, \texttt{IPython}~\cite{Perez:2007ipy}, \texttt{MontePython}~\cite{Audren:2012wb, Brinckmann:2018cvx}, the \texttt{Flexible Atomic Code (FAC, cFAC)} code~\cite{2002ApJ...579L.103G, 2003ApJ...582.1241G, 2008CaJPh..86..675G}, the Mathematica package \texttt{FeynCalc}~\cite{Mertig:1990an, Shtabovenko:2016sxi, Shtabovenko:2020gxv, Shtabovenko:2023idz}, and the Python packages \texttt{Matplotlib}~\cite{Hunter:2007mat}, \texttt{NumPy}~\cite{Harris:2020xlr}, and~\texttt{SciPy}~\cite{Virtanen:2019joe}. 
We also acknowledge the use of the \texttt{AxionLimits} Github repository~\cite{AxionLimits, Caputo:2021eaa}, which was an invaluable resource in verifying existing vector dark matter constraints and extracting most of the bounds from the current literature shown in Fig.~\ref{fig:Vector}.

We acknowledge the Texas Advanced Computing Center (TACC) at The University of Texas at Austin for providing high-performance computing resources that have contributed to the research results reported within this paper.

GM acknowledges support by the Writing Fellowship of the Graduate School of the College of Natural Sciences at the University of Texas at Austin.
NB acknowledges support from the European Union's Horizon Europe research and innovation program under the Marie Sk\l{}odowska-Curie grant agreement no. 101207487 (GWSKY - Mapping the Universe with Gravitational Waves) and by PRD/ARPE 2022 ``Cosmology with Gravitational waves and Large Scale Structure - CosmoGraLSS''.
KB acknowledges support from the National Science Foundation under Grant No.~PHY-2413016.

\end{acknowledgments}


\appendix

\section{Inelastic Scattering Cross Section}
\label{app:DM_crosssection}

In this appendix, we derive the cross sections for the inelastic scattering processes $V\,e^{-}\to\gamma\,e^{-}$ and $A\,e^{-}\to\gamma\,e^{-}$, in which a vector or axial-vector DM particle is converted into a photon, yielding Eqs.~\eqref{eq:sV_scat} and~\eqref{eq:sA_scat} in the main text.

\subsection{Vector case}

The tree-level process $e^{-}\,V\to e^{-}\,\gamma$ receives contributions from the two Feynman diagrams shown in Fig.~\ref{fig:feynman_diagrams}. Each diagram involves a $(-ig_{V\bar{e}e}\gamma^\nu)$ vertex for the DM coupling and a $(-ie\gamma^\mu)$ vertex for the electromagnetic coupling, with $e = \sqrt{4\pi\alpha_{\rm EM}}$.

The corresponding spin- and polarization-averaged squared amplitude is
\begin{widetext}
\begin{align}
\overline{|\mathcal{M}|^2} = \frac{16\pi\,\alpha_{\rm EM}\,g_{V\bar ee}^2}{3\,(s-m_e^2)^2\,(m_e^2+m_V^2-s-t)^2}\bigg[&\, 2\,m_e^8+4\,m_e^6\left(m_V^2-2\,s\right) + m_e^4\left(7\,m_V^4 - 4\,m_V^2(3\,s+2\,t) + 12\,s^2+4\,s\,t+3\,t^2\right) \nonumber \\
+\, m_e^2\bigg(&3\,m_V^6-m_V^4(6\,s+5\,t)+m_V^2(12\,s^2+4\,s\,t+3\,t^2)-8\,s^3-8\,s^2t-2\,s\,t^2-t^3\bigg) \nonumber \\
+\,& s\,(s+t-m_V^2)\left(m_V^4-2\,m_V^2\,s+2\,s^2+2\,s\,t+t^2\right)\bigg]\,,
\label{eq:Msq_vector}
\end{align}
\end{widetext}
where the denominator factors $(s-m_e^2)^2$ and $(m_e^2+m_V^2-s-t)^2\equiv(u-m_e^2)^2$ arise from the squared $s$- and $u$-channel electron propagators, respectively, and we sum over all final-state spins and polarizations while averaging over the $2\times 3=6$ initial-state degrees of freedom (2 electron spins and 3 massive vector polarizations).

The unpolarized differential cross section in the center-of-mass (CM) frame is then given by~\cite{Peskin:1995ev}
\begin{equation}
    \frac{d\sigma}{d\Omega}\bigg|_{\rm CM} =
     \frac{|\vec{p}_f|}{|\vec{p}_i|}\,\frac{\overline{|\mathcal{M}|^2}}{64\,\pi^2\,s}\,,
    \label{eq:dsigma_dOmega}
\end{equation}
where $|\vec{p}_i|$ and $|\vec{p}_f|$ are the magnitudes of the initial and final 3d-momenta in the CM frame, respectively. The ratio of momenta
\begin{equation}
    \frac{|\vec{p}_f|}{|\vec{p}_i|} = \sqrt{\frac{\lambda(s,\,m_e^2,\,0)}{\lambda(s,\,m_e^2,\,m_V^2)}} = \frac{s-m_e^2}{\lambda^{1/2}(s,\,m_e^2,\,m_V^2)} \label{eq:pf_over_pi}
\end{equation}
is conveniently expressed in terms of the K\"all\'en function, $\lambda(a,b,c)=a^2+b^2+c^2-2ab-2ac-2bc$.
 After combing Eqs.~\eqref{eq:dsigma_dOmega} and~\eqref{eq:pf_over_pi} and expressing the Mandelstam variable~$t$ in terms of the CM scattering angle~$\cos\theta$, the total cross section~$\sigma_V(s)$ is obtained by integrating over the full solid angle. The resulting expression is an exact, closed-form function of~$s$, $m_e$, and~$m_V$, which we do not report here for brevity. To extract the non-relativistic (NR) limit relevant at late cosmological times, we simply substitute $s\approx (m_e + m_V)^2 + m_e\,m_V\,v_r^2$, where~$v_r$ is the DM--electron relative velocity, and expand in powers of~$v_r$. The leading-order term scales as~$1/v_r$:
\begin{widetext}
    \begin{align}
    \sigma_{V\,e^{-}\!\to\gamma\,e^{-}} &= \frac{\alpha_{\rm EM}\,g_{V\bar ee}^2}{6\,m_e^2\,v_r}\,\frac{\left(2\,m_e+m_V\right)\!\left(2\,m_e^2+2\,m_e\,m_V+m_V^2\right)}{\left(m_e+m_V\right)^3} +\mathcal{O}(v_r^2)\,,
    \label{eq:sigma_V_full_NR}
\end{align}
\end{widetext}
which, in the limit $m_V\ll m_e$, reduces to Eq.~\eqref{eq:sV_scat}.

\subsection{Axial-vector case}

For the axial-vector model, the calculation proceeds identically to the vector case, with the sole modification being the replacement of the DM--electron vertex $g_{V\bar{e}e}\,\gamma^\nu\to\,g_{A\bar{e}e}\, \gamma^5\gamma^\nu$ in both Feynman diagrams, reflecting the axial-vector structure of the coupling. The spin and polarization averaging is performed with the same $1/6$ factor. Following the same steps, the NR cross section again scales as~$1/v_r$ at leading order:
\begin{widetext}
    \begin{equation}
    \sigma_{A\,e^{-}\!\to\gamma\,e^{-}} = \frac{\alpha_{\rm EM}\,g_{A\bar ee}^2}{6\,m_e^2\,v_r}\,\frac{\left(2\,m_e+m_A\right)\!\left(4\,m_e^2+2\,m_e\,m_A+m_A^2\right)}{\left(m_e+m_A\right)^3}+\mathcal{O}(v_r)\,,
    \label{eq:sigma_A_full_NR}
\end{equation}
\end{widetext}
which, in the limit $m_A\ll m_e$, reduces to Eq.~\eqref{eq:sA_scat}.


\section{Cross Section and Atomic Form Factor for DM-Induced Hydrogen Ionization}
\label{app:hydrogen_ionization_formfactor}

In this appendix, we derive the hydrogen ionization cross sections and the associated form factors~$Q_V(\epsilon)$ and~$Q_A(\epsilon)$ reported in Eqs.~\eqref{eq:sV_Hion},~\eqref{eq:QV} and~\eqref{eq:sA_Hion},~\eqref{eq:QA}, following the formalism of Refs.~\cite{Tan:2021nif,Ibe:2017yqa}. We work in the rest frame of the atom and neglect the nuclear recoil throughout, which is justified in the sub-MeV DM mass range of interest.

\subsection{Ionization cross section}

We consider the process in which a vector DM particle~$V^\mu$ with energy~$\epsilon$ is absorbed by a hydrogen atom, ionizing it. In the NR regime relevant for cold DM at late cosmological times, the massive vector field is dominated by its longitudinal (temporal) component, and the interaction Lagrangian in Eq.~\eqref{eq:Lint_Vscat} reduces to $\mathcal{L}_{\rm int}\approx -g_{V{\bar e}e}V^0{\bar \psi}_e \gamma_0 \psi_e$. With this simplification, the temporal component can then be parametrized as a plane wave:
\begin{equation}
    V^0 = V_\star\,e^{i(\vec{k}\cdot\vec{r}-\omega t)}
\end{equation} 
with wavenumber~$|\vec{k}| = \sqrt{\epsilon^2 - m_V^2}$ and normalization constant~$V_\star$, yielding an interaction Hamiltonian
\begin{equation}
    H_{\rm int} = -g_{V\bar{e}e}\,V_\star\,e^{i\vec{k}\cdot\vec{r}}\,\mathbb{I}\,,
\end{equation}
where~$\bar{\psi}_e\gamma_0\psi_e = \psi_e^\dagger(\gamma^0)^2\psi_e = \psi_e^\dagger\,\mathbb{I}\,\psi_e$. 

Applying Fermi's golden rule, the total rate at which the atom absorbs the DM particle and ejects an electron into the continuum is
\begin{align}
    W &= 2\pi\sum_{bc}|\langle b|H_{\rm int}|c\rangle|^2 \nonumber \\
    &=2\pi\,g_{V\bar{e}e}^2\,V_*^2\sum_{bc}\left|\int e^{i\vec{k}\cdot\vec{r}}\,\psi_{e,b}^\dagger\,\mathbb{I}\,\psi_{e,c}\,d^3r\right|^2,
    \label{eq:W_vector}
\end{align}
where the sum runs over all initial bound states~$b$ and final continuum states~$c$, and energy conservation~$\epsilon_c = \epsilon_b + \epsilon$ is left implicit. The ionization cross section is obtained by dividing the incoming particle flux,
$j_V = 2|\vec{k}|\,V_\star^2 = 2\,\epsilon\,v\,V_\star^2$~\cite{Tan:2021nif}, yielding
\begin{align}
    \sigma_{V+H\rightarrow e^{-}+p^{+}} = \frac{W}{j_V} &= g_{V\bar{e}e}^2\,\frac{\pi}{\epsilon\,v}\sum_{bc}\left|\int e^{i\vec{k}\cdot\vec{r}}\,\psi_{e,b}^\dagger\,\mathbb{I}\,\psi_{e,c}\,d^3r\right|^2 \nonumber \\
    &\equiv g_{V\bar{e}e}^2\,\frac{a_0^2}{v}\,Q_V(\epsilon)\,,
    \label{eq:sigma_V_app}
\end{align}
where~$a_0$ is the Bohr radius and the dimensionless form factor~$Q_V(\epsilon)$ is defined in Eq.~\eqref{eq:QV}. Since the derivation is carried out in the rest frame of the atom, the velocity~$v$ appearing in Eq.~\eqref{eq:sigma_V_app} corresponds, in the NR limit, to the DM--hydrogen relative velocity~$v_r$ used in the main text.
 
For the axial-vector model, the interaction Lagrangian~$\mathcal{L}_{\rm int} = g_{A\bar{e}e}\,A_\mu\,\bar{\psi}_e\gamma^5\gamma^\mu\psi_e$ leads, in the same NR limit, to a Hamiltonian proportional to~$\gamma^0\gamma^5\gamma^0 = \gamma^5$. The derivation proceeds identically to the vector case, with the replacement~$\mathbb{I}\to\gamma^5$ in the matrix element, yielding the form factor~$Q_A(\epsilon)$ defined in Eq.~\eqref{eq:QA}.

\subsection{Evaluation of the atomic form factor}

To evaluate the form factors numerically, we decompose the electron wavefunctions using the two-component spherical
spinors $\Omega_m^\kappa(\theta,\phi)$ as~\cite{Tan:2021nif,Ibe:2017yqa}
\begin{equation}
    \psi_e(\vec{r}) = \frac{1}{r}\begin{pmatrix} f_\epsilon^\kappa(r)\,\Omega_m^\kappa(\theta,\phi) \\ i\,g_\epsilon^\kappa(r)\,\Omega_m^{-\kappa}(\theta,\phi) \end{pmatrix}\,,
    \label{eq:dirac_spinor}
\end{equation}
where~$f$ and~$g$ are the radial components of the electron wavefunction, and the quantum number~$\kappa \equiv (j+1/2)(-1)^{l+1}$ encodes both the total and orbital angular momenta.

Substituting this decomposition into the matrix element in Eq.~\eqref{eq:W_vector}, choosing the reference frame~$\vec{k} = k\,\hat{e}_z$ without loss of generality, and expanding the plane wave in spherical Bessel functions, $e^{ikz} = \sum_L (-i)^L(2L+1)\,j_L(kr)\,P_L(\cos\theta)$, the angular integrations can be carried out analytically using the Wigner-Eckart theorem. The three-dimensional integral then reduces to a multipole sum of one-dimensional radial integrals:\footnote{See Ref.~\cite{Tan:2021nif} for a detailed derivation of this reduction in the scalar DM case, which we generalize here to the vector and axial-vector interactions.}
\begin{widetext}
\begin{equation}
    \langle b|| \mathcal{O}^{V/A}_L||c\rangle\equiv\int e^{i\vec{k}\cdot\vec{r}}\,\psi_{e,b}^\dagger\,\Gamma_{V/A}\,\psi_{e,c}\,d^3r = \sum_{L=0}^\infty (-i)^L\,(2L+1)\,\langle \kappa_b\|C_{L}\|\kappa_c\rangle_{V/A}\;\int_0^\infty \mathcal{R}^{V/A}_{\epsilon_{b,c}\,\kappa_{b,c}}(r)\, j_L(kr)\, dr\,,
    \label{eq:multipole_decomp}
\end{equation}
\end{widetext}
where~$\Gamma_{V/A}$,~$\mathcal{R}^{V/A}_{\epsilon_{b,c}\,\kappa_{b,c}}$, and~$\langle \kappa_b\|C_{L}\|\kappa_c\rangle_{V/A}$ denote the Dirac matrix, the radial integrand, and the angular coefficient entering the multipole decomposition, respectively, each of which takes a different form for the vector and axial-vector interactions.
Namely, the radial integrands are given by
\begin{align}
    \mathcal{R}^{V}_{\epsilon_b,\,\epsilon_c} &= f_{\epsilon_b}^{\kappa_b}\,f_{\epsilon_c}^{\kappa_c} + g_{\epsilon_b}^{\kappa_b}\,g_{\epsilon_c}^{\kappa_c}\,, \\
    \mathcal{R}^{A}_{\epsilon_b,\,\epsilon_c} &= f_{\epsilon_b}^{\kappa_b}\,g_{\epsilon_c}^{\kappa_c} - g_{\epsilon_b}^{\kappa_b}\,f_{\epsilon_c}^{\kappa_c}\,,
\end{align}
with~$\Gamma_V = \mathbb{I}$ and $\Gamma_A = \gamma_5$, while the angular coefficient reads
\begin{align}
    \langle \kappa_b\|C_{L}\|\kappa_c\rangle_{V/A} &= (-1)^{j_b-1/2}\sqrt{(2j_b+1)(2\tilde{j}_c+1)} \nonumber \\
    &\times\begin{pmatrix} j_b & \tilde{j}_c & L \\ -\tfrac{1}{2} & \tfrac{1}{2} & 0 \end{pmatrix}\Pi(l_b+L+\tilde{l}_c)\,,
\end{align}
where $\Pi(x)$ enforces parity selection, returning~1 if~$x$ is even and~$0$ otherwise. In the vector case, the continuum quantum numbers enter directly~$(\tilde{\kappa}_c,\,\tilde{j}_c,\,\tilde{l}_c) = (\kappa_c,\,j_c,\,l_c)$, while for the axial-vector case they are determined by~$-\kappa_c$, reflecting the parity flip induced by~$\gamma_5$ on the lower spinor component. The same chiral structure is responsible for the cross terms between upper and lower radial components appearing in~$\mathcal{R}^A$, in contrast to the diagonal combination in~$\mathcal{R}^V$. 

Combining Eqs.~\eqref{eq:sigma_V_app} and~\eqref{eq:multipole_decomp}, the atomic form factors can be written in the final form
\begin{equation}
    Q_{V/A}(\epsilon) = \frac{\pi}{\epsilon\,a_0^2}\sum_{bc}\sum_{L=0}^\infty(2L+1)\left|\,\langle b\|\mathcal{O}^{V/A}_L\| c\rangle\right|^2.
    \label{eq:Q_final}
\end{equation}

The bound-state electron wavefunctions entering the radial integrals are normalized to unity (dimensionless), while the continuum states are normalized to~$\delta(\epsilon_c - \epsilon_c^\prime)$, giving them dimensions of~$1/\sqrt{\text{energy}}$. Both are computed numerically using the \texttt{Flexible Atomic Code (FAC, cFAC)} code~\cite{2002ApJ...579L.103G, 2003ApJ...582.1241G, 2008CaJPh..86..675G}.

The multipole sum in Eq.~\eqref{eq:Q_final} converges very rapidly in practice, and we truncate it at~$L=3$. As discussed in the main text, the relative importance of the leading multipoles differs between the two interaction types. In the vector case, the monopole ($L=0$) contribution is strongly suppressed, because the orthogonality of the bound and continuum radial wavefunctions implies~$\int(f_b\,f_c + g_b\,g_c)\,dr = 0$, and in the NR limit of interest, with $ kr\ll 1$, the factor~$j_0(kr)\approx 1$, making the monopole integral nearly vanish~\cite{Tan:2021nif}. The dominant contribution to~$Q_V$ therefore arises from the dipole ($L=1$) term, in close analogy with standard photoionization. In the axial-vector case, the cross terms in~$\mathcal{R}^A$ are not constrained by this orthogonality condition, so the monopole is not suppressed and provides the leading contribution to~$Q_A$. This difference accounts for the sharper peak and reduced numerical noise of the axial-vector form factor relative to the vector case, as shown in Fig.~\ref{fig:Q_form_factor}.

\begin{figure}
    \centering
    \includegraphics[width=\linewidth]{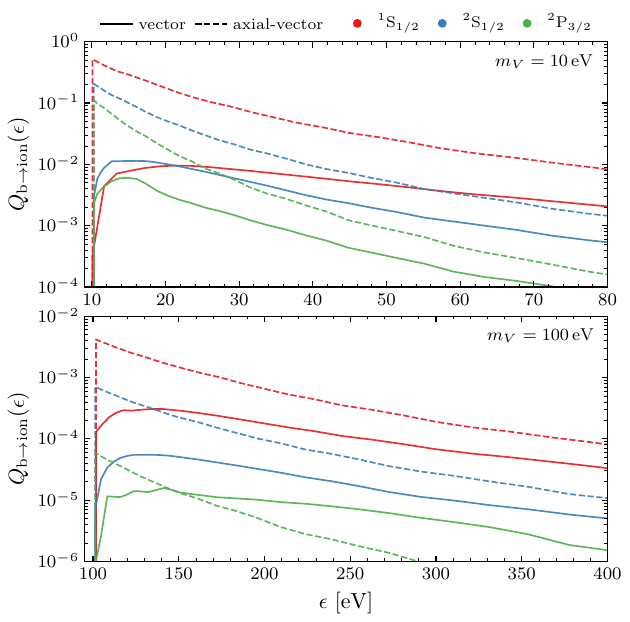}
    \caption{{\it Top panel}: Hydrogen atomic form factor~$Q$ as a function of the incoming DM energy~$\epsilon$ for $m_{\rm DM}=10\,$eV, decomposed into contributions from the~$^1 S_{1/2}$~(\textcolor{tabred}{red}),~$^2 S_{1/2}$~(\textcolor{tabblue}{blue}), and~$^2 P_{3/2}$~(\textcolor{tabgreen}{green}) bound states. Solid curves correspond to the vector interaction, while dashed curves show the axial-vector case. {\it Bottom panel}: Same as the {\it Top} but for $m_{\rm DM} = 100\,$eV. For the higher DM mass, the ground state~$^1S_{1/2}$ clearly dominates in both scenarios, while at~$m_{\rm DM} = 10\,$eV, the hierarchy among bound states weakens, particularly for the vector interaction.}
    \label{fig:A2}
\end{figure}

\subsection{Contributions from individual bound states}

So far, the form factors in Eq.~\eqref{eq:Q_final} have been written as sums over all initial bound states~$b$. However, as discussed in the main text, in our analysis we retain only the contribution from the hydrogen ground state~$^1 S_{1/2}$, since for DM masses well above the binding energies, $m_{\rm DM} \gg \mathcal{O}(10\,\text{eV})$, the most tightly bound state dominates the form factor---a hierarchy that mirrors the well-known behavior of standard photoionization~\cite{bethe:photoionization, sobelman:photoionization}.

We illustrate this explictly in Fig.~\ref{fig:A2}, which compares the individual contributions from the~$^1 S_{1/2}$, $^2 S_{1/2}$, and~$^2 P_{3/2}$ states for both the vector and axial-vector interactions at two representative DM masses.  For~$m_{\rm DM} = 100\,$eV the~$^1 S_{1/2}$ state clearly dominates in both scenarios, while at~$m_{\rm DM} = 10\,$eV----which is closer to the individual ionization thresholds---the hierarchy weakens considerably, with the~$^2 S_{1/2}$ contribution becoming comparable to, and in the vector case even slightly exceeding, that of the~~$^1 S_{1/2}$ state.

It is partly for this reason that we restrict our analysis to DM masses~$m_{\rm DM} \geq 100\,$eV, where the ground-state approximation is well justified. We note, however, that even without this restriction, the contribution of excited states to the cosmological energy injection would remain negligible in practice, since outside the narrow window of recombination, essentially all hydrogen atoms reside in the ground state.


\section{Derivation of the DM-electron Absorption Rate}
\label{app:DM_absorption_rate}

In this appendix, we derive the thermally averaged cross sections entering the energy injection rates for the two DM conversion processes discussed in Sec.~\ref{sec:energy_injection}.
 
\subsection{DM conversion via inelastic scattering}
 
The interaction rate per unit volume for the inelastic process $(V/A)\,e^-\to\gamma\,e^-$ is obtained by integrating the cross section over the phase-space distributions of the incoming DM and electron. In the NR limit, relevant at late cosmological times, both species follow Maxwell--Boltzmann distributions, and we can write
\begin{align}
    \langle \sigma v \rangle = \frac{1}{(\bar{v}_e\,\bar{v}_{\rm DM})^3}&\int\frac{d^3v_e\,d^3v_{\rm DM}}{(2\pi)^3}\,e^{-v_e^2/(2\bar{v}_e^2)} \nonumber \\
    &\times \,e^{-v_{\rm DM}^2/(2\bar{v}_{\rm DM}^2)}\;\sigma(v_r)\,v_r\,,
    \label{eq:rate_compton}
\end{align}
where~$n_e$ and~$n_{\rm DM}$ are the free-electron and DM number densities, $\bar{v}_j \equiv \sqrt{T_j/m_j}$ are the thermal velocity dispersions with $j\in\{e,\,{\rm DM}\}$,~$v_r = |\vec{v}_{\rm DM}-\vec{v}_e|$ is the relative velocity, and we have dropped the subscript denoting the interaction type for clarity.

Since the cross section depends only on~$v_r$, it is convenient to change variables from~$\{\vec{v}_e,\,\vec{v}_{\rm DM}\}$ to $\{\vec{v}_{\rm m},\,\vec{v}_r\}$, where~$\vec{v}_{\rm m}$ is a weighted mean velocity defined such that the two Gaussians in Eq.~\eqref{eq:rate_compton} factorize, namely:
\begin{equation}
\begin{gathered}
    \vec{v}_{\rm m} = \frac{\bar{v}_e^2\,\vec{v}_e + \bar{v}_{\rm DM}^2\,\vec{v}_{\rm DM}}{\bar{v}_e^2+\bar{v}_{\rm DM}^2}\,,\\[6pt]
    \bar{v}_{\rm m}^2 = \frac{\bar{v}_e^2\,\bar{v}_{\rm DM}^2}{\bar{v}_e^2+\bar{v}_{\rm DM}^2}\,,\qquad
    \bar{v}_r^2 = \bar{v}_e^2+\bar{v}_{\rm DM}^2\,.
\end{gathered}
\end{equation}
The integral over~$\vec{v}_{\rm m}$ then evaluates to unity, leaving only the integration over the relative velocity:
\begin{equation}
    \langle\sigma v\rangle = \,\frac{1}{(\bar{v}_r)^3}\int\frac{d^3v_r}{(2\pi)^{3/2}}\,e^{-v_r^2/(2\bar{v}_r^2)}\;\sigma(v_r)\,v_r\,.
    \label{eq:rate_vr}
\end{equation}
As shown in Appendix~\ref{app:DM_crosssection}, the NR cross sections for inelastic scattering, Eqs.~\eqref{eq:sV_scat} and~\eqref{eq:sA_scat}, scale as~$\sigma = \sigma_0/v_r$, where~$\sigma_0$ is a velocity-independent prefactor. Since the product~$\sigma(v_r)\,v_r = \sigma_0$ is then constant, it factors out of the remaining integral in Eq.~\eqref{eq:rate_vr}, which reduces to the normalization of the Gaussian. The thermally averaged cross section is therefore simply
\begin{equation}
\langle\sigma v\rangle_{(V/A)\,e^-\to\gamma\,e^-} = \sigma_0 = \sigma_{(V/A)\,e^-\to\gamma\,e^-}\,v_r\,.
\label{eq:sigmav_compton}
\end{equation}
 Substituting the explicit expressions for the cross sections gives~$\langle\sigma v\rangle = \alpha_{\rm EM}\,g_{V\bar{e}e}^2(m_V+2\,m_e)/[6\,m_e^2(2\,m_V+m_e)]$ for the vector case, Eq.~\eqref{eq:sV_scat}, and~$\langle\sigma v\rangle = \alpha_{\rm EM}\,g_{A\bar{e}e}^2(m_A+4\,m_e)/[6\,m_e^2(2\,m_A+m_e)]$ for the axial-vector case, Eq.~\eqref{eq:sA_scat}.
 
\subsection{DM absorption via hydrogen ionization}
 
For hydrogen ionization, the thermal averaging proceeds analogously to the inelastic scattering case, with the free-electron number density~$n_e$ replaced by the neutral hydrogen density~$n_{\rm HI}$, and the electron velocity now referring to that of the bound electron within the hydrogen atom. The cross sections, Eqs.~\eqref{eq:sV_Hion} and~\eqref{eq:sA_Hion}, again scale as~$1/v_r$, but unlike the inelastic case, the prefactor is not strictly constant, as it depends on the total energy~$\epsilon$ of the incoming DM particle through the atomic form factor~$Q_{V/A}(\epsilon)$.
 
In principle, this energy dependence must be retained inside the phase-space integral. However, as shown in Fig.~\ref{fig:Q_form_factor}, for~$m_{\rm DM}\gg E_{H,\,\rm ion}$, the form factor is generically peaked at energies of order the DM mass, with the thermal averaging over the cold DM phase-space distribution that further weights the contribution toward~$\epsilon\approx m_{\rm DM}$. We can therefore approximate~$Q_{V/A}(\epsilon)\approx Q_{V/A}(m_{\rm DM})$ and factor it out of the integral. This reduces the cross section to the same~$\sigma_0/v_r$ form encountered in the inelastic scattering case, with a velocity-independent prefactor~$\sigma_0 = g_{(V/A)\,\bar{e}e}^2\,Q_{V/A}(m_{\rm DM})\,a_0^2$. The result of Eq.~\eqref{eq:sigmav_compton} then directly applies, giving
\begin{equation}
    \langle\sigma v\rangle_{(V/A)\,H\to p^+\,e^-} \approx g_{(V/A)\,\bar{e}e}^2\,a_0^2\,Q_{V/A}(m_{\rm DM})\,.
\end{equation}


\section{DM Absorption Implementation in \texttt{DarkHistory}}
\label{app:DM_in_DarkHistory}

\begin{figure*}
    \centering
    \includegraphics[width=\linewidth]{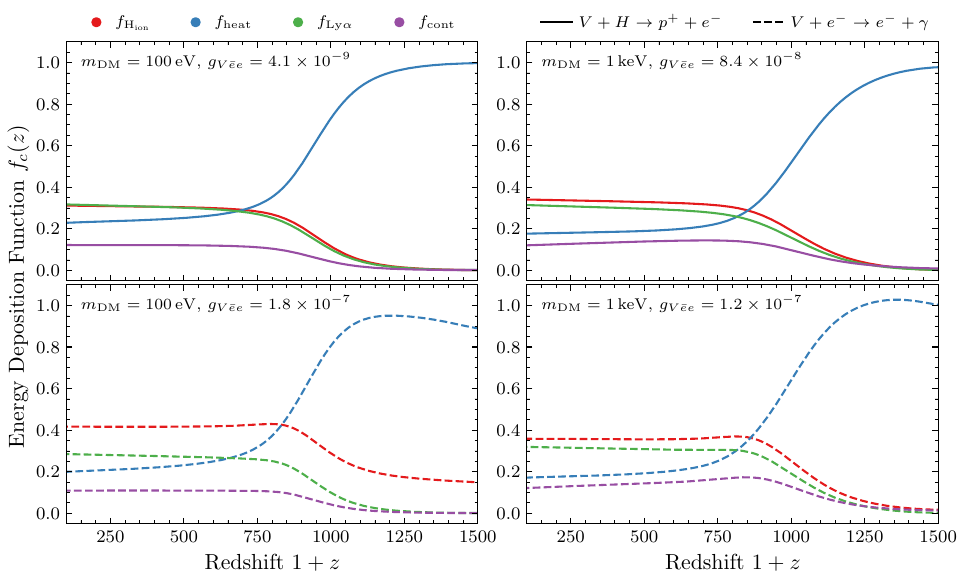}
    \caption{Energy deposition functions~$f_c(z)$ as a function of redshift for the vector DM scenario, computed with the modified version of \texttt{DarkHistory v2.0} described in this appendix. The {\it Left} and {\it Right panels} correspond to $m_{\rm DM} = 100\,$eV and $m_{\rm DM} = 1\,$keV, respectively, with the DM--electron coupling~$g_{V\bar e e}$ fixed at the corresponding {\it Planck} 95\% C.L. upper limit. In each panel, the {\it top} rows show the deposition functions for DM absorption through hydrogen ionization (solid lines), while the {\it bottom} rows correspond to DM conversion into a photon via Compton-like scattering (dashed lines). The four deposition channels --- hydrogen ionization, heating, Lyman-$\alpha$ excitation, and continuum photon production --- are displayed in the colors indicated in the legend.}
    \label{fig:A1}
\end{figure*}

We compute the energy deposition functions $f_c(z)$ using \texttt{DarkHistory v2.0}~\cite{Liu:2019bbm, Liu:2023fgu}, which we modify to include the DM conversion processes discussed in Sec.~\ref{sec:energy_injection}. In this appendix, we summarize the main modifications to the code.

For each conversion channel, we add to the injection module the corresponding energy injection rates presented in Sec.~\ref{sec:energy_injection}, and specify the spectrum of SM particles produced per absorption event.

For DM inelastic scattering with a free-electron, each event injects a single photon and a single electron into the plasma, whose energies are fixed by momentum conservation. In the center-of-mass frame, the kinematics of the  $(V/A)\,e^- \to \gamma\,e^-$ process gives
\begin{equation}
    \sqrt{m_e^2 + p_e^2} + \sqrt{m_{\rm DM}^2 + p_e^2} = p_\gamma + \sqrt{m_e^2 + p_\gamma^2}\,,
\end{equation}
where we used~$|\vec{k}| = |\vec{p}_e|$ and $|\vec{p}_\gamma| = |\vec{p}_e^{\, \prime}|$, with $\vec{p}_e$ and $\vec{p}_e^{\, \prime}$ representing the three-momentum  of the incoming and outgoing electron, respectively. Expanding in the NR limit $p^2_e/\left(2\,m_{\rm DM}\right),\,p^2_e/\left(2\,m_{\rm e}\right) \ll 1$, relevant at late times, we obtain the photon energy and outgoing electron kinetic energy:
\begin{equation}
   p_\gamma \approx \frac{m_{\rm DM}(m_{\rm DM} + 2m_e)}{2(m_{\rm DM} + m_e)}\,, \qquad K_e \approx \frac{p_\gamma^2}{2m_e}\, 
\end{equation}
from which the total injected energy of Eq.~\eqref{eq:Einj_scat} immediately follows. 

For DM absorption through hydrogen ionization, each $(V/A)\,H \to p^+\,e^-$ event instead produces a single electron that carries away essentially all of the DM rest-mass energy as kinetic energy, $K_e\approx m_{\rm DM}$.
The recoiling proton receives a negligible fraction of the injected energy and is therefore not included in the injection spectrum.

Finally, by default \texttt{DarkHistory} assumes that injected electrons are accompanied by a positron, and that after they have lost all of their kinetic energy, they form positronium and annihilate promptly, producing a gamma-ray spectrum, which is added to the propagating photon population. Since the DM conversion processes considered here produce only single electrons, we disable this positronium annihilation step to avoid introducing a spurious source of photons into the calculation.

As a representative example, in Fig.~\ref{fig:A1} we show the energy deposition functions $f_c(z)$ computed for the vector DM scenario at two benchmark masses, $m_{\rm DM} = 100\,$eV and $1\,$keV, with the DM-electron coupling fixed at the corresponding \textit{Planck} 95\% C.L. upper limits derived in this work. The axial-vector case yields effectively equivalent results and is not shown separately. 

In the figure, we display the four primary deposition channels---hydrogen ionization, Lyman-$\alpha$ excitations, heating of the intergalactic
medium, and low-energy continuum
photons production---with the heating channel dominating the energy budget at high redshifts, while hydrogen ionization and Lyman-$\alpha$ excitation take over at later times.

\bibliography{references.bib}

\end{document}